\documentclass[12pt, preprint]{aastex}
\usepackage{amsmath}
\usepackage{graphicx}
\usepackage{natbib}

\usepackage[colorlinks=true,citecolor=blue,breaklinks=true,linktocpage=true]{hyperref}
\bibpunct{(}{)}{;}{a}{}{,} 
\usepackage{xspace}

\usepackage{xparse} 
\usepackage{multirow}

\shorttitle{Sausage modes in nonuniform loops}
\shortauthors{Chen et al.}

\begin{document}

\title{STANDING SAUSAGE MODES IN NONUNIFORM MAGNETIC TUBES:
   AN INVERSION SCHEME FOR INFERRING FLARE LOOP PARAMETERS}

\author{Shao-Xia Chen\altaffilmark{1}}
\author{Bo Li\altaffilmark{1}}
\email{bbl@sdu.edu.cn}
\author{Ming Xiong\altaffilmark{2}}
\author{Hui Yu\altaffilmark{1}}
\and
\author{Ming-Zhe Guo\altaffilmark{1}}

\altaffiltext{1}{Shandong Provincial Key Laboratory of Optical Astronomy and Solar-Terrestrial Environment, Institute of Space Sciences, Shandong University,Weihai, 264209, China}
\altaffiltext{2}{National Space Science Center, CAS, 100190 Beijing, China}

\begin{abstract}
Standing sausage modes in flare loops are important for interpreting quasi-periodic pulsations (QPPs)
    in solar flare lightcurves.
We propose an inversion scheme that {consistently uses their periods $P$ and damping times $\tau$ 
    to diagnose flare loop parameters}.
We derive a generic dispersion relation governing linear sausage waves in
    pressure-less straight tubes, for which the transverse density
    inhomogeneity takes place in a layer of arbitrary width $l$ and is of arbitrary form.
We find that $P$ and $\tau$ depend on the combination of $[R/v_{\rm Ai}, L/R, l/R, \rho_{\rm i}/\rho_{\rm e}]$, where
    $R$ is the loop radius, $L$ is the looplength, $v_{\rm Ai}$ is the internal Alfv\'en speed,
    and $\rho_{\rm i}/\rho_{\rm e}$ is the density contrast.
For all the density profiles examined, $P$ and $\tau$ experience saturation when $L/R \gg 1$,
    yielding an inversion curve in the $[R/v_{\rm Ai}, l/R, \rho_{\rm i}/\rho_{\rm e}]$ space
    with a specific density profile when $L/R$ is sufficiently large.
{When applied to a spatially unresolved QPP event, the scheme yields that
    $R/v_{\rm Ai}$ is} the best constrained,
    whereas $l/R$ corresponds to the other extreme.
For spatially resolved QPPs, while $L/R \gg 1$ cannot be assumed beforehand, an inversion curve remains possible
    due to additional geometrical constraints.
When a spatially resolved QPP event involves another mode, as is the case for a recent event,
    the {full set of} $[v_{\rm Ai}, l, \rho_{\rm i}/\rho_{\rm e}]$ can be inferred.
We conclude that the proposed scheme provides a useful tool for {magneto-seismologically} exploiting QPPs.
\end{abstract}

\section{INTRODUCTION}
\label{sec_intro}

{The original ideas that laid the foundation 
    for the field of solar magneto-seismology (SMS)}
    were put forward in the 1970s 
    (\citeauthor{1970PASJ...22..341U}~\citeyear{1970PASJ...22..341U},
    \citeauthor{1970A&A.....9..159R}~\citeyear{1970A&A.....9..159R},
    \citeauthor{1975IGAFS..37....3Z}~\citeyear{1975IGAFS..37....3Z},
    see also \citeauthor{1984ApJ...279..857R}~\citeyear{1984ApJ...279..857R}).
{However, this} field  flourished only
    after a rich variety of low-frequency Magnetohydrodynamic (MHD) waves and oscillations
    were identified
    with the advent of the TRACE, SOHO, Hinode, and SDO satellites
    (for recent reviews, see
    \citeauthor{2007SoPh..246....3B}~\citeyear{2007SoPh..246....3B},
    \citeauthor{2012RSPTA.370.3193D}~\citeyear{2012RSPTA.370.3193D},
    {\citeauthor{2013SSRv..175....1M}~\citeyear{2013SSRv..175....1M}};
    and also~\citeauthor{2007SoPh..246....1B}~\citeyear{2007SoPh..246....1B},
    \citeauthor{2009SSRv..149....1N}~\citeyear{2009SSRv..149....1N},
    \citeauthor{2011SSRv..158..167E}~\citeyear{2011SSRv..158..167E} for three recent topical issues).
It is also indispensable to refine the theoretical understanding of the collective
    wave modes supported by magnetized tubes,
    thereby enabling one to employ the measured wave properties to
    infer the solar atmospheric parameters that
    are difficult to measure
    directly~\citep[e.g.,][]{2000SoPh..193..139R, 2005LRSP....2....3N}.
{Regarding its applications to the solar corona, SMS} can offer such key
    information as
    the magnetic field strength in coronal loops
    \citep[e.g.,][]{2001A&A...372L..53N,2008A&A...489L..49E, 2008A&A...482L...9O, 2012A&A...537A..49W}
    and above streamer stalks~\citep{2010ApJ...714..644C,2011ApJ...728..147C},
    the magnitude of field-aligned loop flows~\citep{2013ApJ...767..169L, 2014SoPh..289.1663C},
    the temperature of loop plasmas~\citep[e.g.,][]{2009ApJ...706L..76M},
    the coronal effective adiabatic index~\citep{2011ApJ...727L..32V},
    as well as the {longitudinal~\citep{2008A&A...486.1015V,2009SSRv..149....3A,2012ApJ...748..110L} 
    and transverse structuring}~\citep[e.g.,][]{2007A&A...463..333A,2008A&A...484..851G,2015ApJ...799..221Y}.
{In addition,
    SMS applications with torsional Alfv\'en waves have proven invaluable in inferring the magnetic field structure
    at chromospheric heights \citep{2009Sci...323.1582J,2011ApJ...740L..46F}.
Likewise, \citet{2012ApJ...748..110L} demonstrated the potential of using longitudinal waves to infer the
    longitudinal variation of density and magnetic field strength in chromospheric waveguides.}

{Magneto-seismological} applications with standing kink modes (with azimuthal wavenumber $m=1$)
    have been a common practice since their
    detection {with TRACE~\citep{1999ApJ...520..880A}}.
Kink oscillations tend to experience substantial
    {damping~\citep[e.g.,][and references therein]{2002ApJ...576L.153O, 2009SSRv..149..199R, 2013A&A...552A.138V}},
    which is usually interpreted in terms of resonant absorption
    (\citeauthor{2002ApJ...577..475R}~\citeyear{2002ApJ...577..475R},
     \citeauthor{2002A&A...394L..39G}~\citeyear{2002A&A...394L..39G},
    also~\citeauthor{1988JGR....93.5423H}~\citeyear{1988JGR....93.5423H},
    {and the comprehensive review by~\citeauthor{2011SSRv..158..289G}~\citeyear{2011SSRv..158..289G}}).
{With} this interpretation, \citet{2002ApJ...577..475R} and
\citet{2002A&A...394L..39G} suggested
    that the measured period $P$ and damping time $\tau$ can be used to
    infer the lengthscale $l$ of the density inhomogeneity across coronal loops
    in units of loop radius $R$.
For this purpose, the largely unknown transverse density distribution was shown to be important,
    since its formulation has a considerable impact on $P$ and $\tau$~\citep{2013ApJ...777..158S,2014ApJ...781..111S}.

While kink modes have attracted much attention,
    sausage modes (with $m=0$) are equally important in SMS.
In fact, sausage modes    
    are even more important from the standpoint of solar atmospheric heating
    given their stronger compressibility and ubiquity in the lower solar
    atmosphere
    \citep{2011ApJ...729L..18M, 2012NatCo...3E1315M, 2014ApJ...791...61F, 2014A&A...563A..12D,2015ApJ...806..132G,2015A&A...579A..73M}.
In addition, sausage modes are important for interpreting 
    quasi-periodic pulsations (QPPs)
    in the lightcurves of solar flares~\citep[see][for a recent review]{2009SSRv..149..119N}.
Two distinct regimes are known to exist, depending on
    the axial wavenumber $k$ along flare loops~\citep{2005LRSP....2....3N}.
The trapped regime results when $k$ exceeds some critical value $k_{\rm c}$,
    where the energy of sausage modes is well confined to magnetic tubes.
{When} $k<k_{\rm c}$,
    the leaky regime arises and sausage modes experience apparent temporal damping by radiating their energy
    into the surrounding fluid~\citep{1982SoPh...75....3S,1986SoPh..103..277C}.
It is known that $k_{\rm c}$ depends sensitively on the density contrast between
    loops and their surroundings~\citep[e.g.,][]{2007AstL...33..706K}. 
In addition,
    both eigen-mode analyses~\citep{2007AstL...33..706K,2014ApJ...781...92V}
    and numerical simulations from an initial-value-problem perspective~\citep{2012ApJ...761..134N,2015SoPh..tmp..118C}
    {indicated} that the period $P$ of sausage modes increases smoothly with decreasing $k$
    (or equivalently with increasing looplength $L$ given that $k = \pi/L$ for fundamental modes)
    until reaching some saturation value $P_{\rm s}$ for sufficiently thin loops ($R/L \ll 1$).
Likewise, identically infinite in the trapped regime for ideal MHD fluids,
    the attenuation time $\tau$ decreases
    with decreasing $k$ before experiencing saturation at $\tau_{\rm s}$ when $R/L \ll 1$.

{Magneto-seismological applications of sausage modes
    are possible due to their dependence on atmospheric parameters
    \citep{2012ApJ...748..110L,2012IAUS..286..437L}.
The practice based on the measured period and damping time     
    can be illustrated by the study} presented in~\citet{2007AstL...33..706K} where
    a step-function (top-hat) form {was} adopted for the transverse density distribution.
{The saturation values, $P_{\rm s}$ and $\tau_{\rm s}/P_{\rm s}$, for large density contrasts are
    approximately} 
\begin{eqnarray}
\label{eq_step_PQ}
P_{\rm s} \approx
    2.62 \frac{R}{v_{\rm Ai}}, \hspace{0.2cm}
\frac{\tau_{\rm s}}{P_{\rm s}} \approx
    \frac{1}{\pi^2}\frac{\rho_{\rm i}}{\rho_{\rm e}} . 
\end{eqnarray}
As an example, \citet{2007AstL...33..706K} examined the {QPP} in the radio emissions
    reported in~\citet{1973SoPh...32..485M},
    where $P$ and $\tau/P$
    were found to be $\sim 4.3$ secs and $\sim 10$, respectively.
With the damping attributed to wave leakage, Eq.~(\ref{eq_step_PQ})
    then {yields a density contrast $\rho_{\rm i}/\rho_{\rm e}$ of}
    $\sim 100$, and
    a transverse Alfv\'en transit time $R/v_{\rm Ai}$ of $\sim 1.64$ secs,
    provided that the flaring loop in question is sufficiently thin.
However, the dependence of $P$ and $\tau$ on {loop parameters is}
    substantially more involved if one goes a step closer to reality
    by replacing a step-function {density profile} with a smooth one.
Even for thin loops where neither $P$ nor $\tau$ depends on looplength,
    both $P_{\rm s}$~\citep{2012ApJ...761..134N} and $\tau_{\rm s}$~\citep{2015SoPh..tmp..118C}
    may be sensitive to the steepness, or equivalently the lengthscale, of the transverse density distribution.
{In mathematical terms, this means} that for a given density profile,
\begin{eqnarray}
\label{eq_FG}
\begin{array}{rcl}
&& P_{\rm s} = \displaystyle\frac{R}{v_{\rm Ai}}
    F\left(\displaystyle\frac{l}{R}, \displaystyle\frac{\rho_{\rm i}}{\rho_{\rm e}} \right), \\ [0.4cm]
&& \displaystyle\frac{\tau_{\rm s}}{P_{\rm s}} =
    G\left(\displaystyle\frac{l}{R}, \displaystyle\frac{\rho_{\rm i}}{\rho_{\rm e}} \right).
\end{array}
\end{eqnarray}
{When} only $P_{\rm s}$ and $\tau_{\rm s}$ are known as is the case for measurements without imaging capabilities,
    the appearance of $l/R$ no longer
    allows a unique pair of $[\rho_{\rm i}/\rho_{\rm e}, R/v_{\rm Ai}]$ to be deduced.
Despite this, one can still constrain {the combination} $[R/v_{\rm Ai}, \rho_{\rm i}/\rho_{\rm e}, l/R]$
    by developing a scheme in much the same way that kink modes were employed~\citep{2007A&A...463..333A, 2008A&A...484..851G,2014ApJ...781..111S},
    the only difference being that the transverse Alfv\'en time $R/v_{\rm Ai}$
    replaces the longitudinal one $L/v_{\rm Ai}$.

The present study aims to develop the aforementioned scheme employing measured periods and damping times
    of standing sausage modes.
An essential ingredient will be to establish the functions $F$ and $G$ in Eq.~(\ref{eq_FG}).
To this end, we will derive an analytical dispersion relation (DR)
    governing linear sausage waves
    hosted by magnetized tubes with a rather general transverse density distribution.
The only requirement here is that this density distribution can be decomposed into a uniform cord,
    a uniform external medium, and a transition layer connecting the two.
However, the density distribution in the transition layer is allowed to be arbitrary, thereby
    making the DR applicable to a rich variety of density profiles.
We note that this kind of density profiles has been extensively adopted in kink mode
    studies~\citep[e.g.,][and references therein]{2014ApJ...781..111S}.
We further note that developing an analytical DR is important in its own right.
Apart from the step-function profile~\citep[e.g.,][]{1982SoPh...75....3S,1986SoPh..103..277C},
    analytical DRs in the cylindrical case are available only
    for a limited set of density profiles~\citep{1986NASCP2449..347E,1988A&A...192..343E,2014A&A...572A..60L}.

This manuscript is organized as follows.
Section~\ref{sec_model} presents 
    {the derivation of the DR and 
    our solution method}.
A parameter study is {presented} in Sect.~\ref{sec_NumRes}
    to examine how the period and damping time of sausage modes depend on {tube parameters},
    thereby establishing our numerical scheme for inverting {measurements
    of spatially unresolved QPPs}.
An {extension to spatially resolved QPPs
    is then }given in Sect.~\ref{sec_spatial_res}.
Finally, Sect.~\ref{sec_conc} closes this manuscript with our summary
    and some concluding remarks.

\section{MATHEMATICAL FORMULATION}
\label{sec_model}
\subsection{Description for the Equilibrium Tube}
\label{sec_sub_equilibrium}
We consider sausage waves in a structured corona modeled by a density-enhanced cylinder with
   radius $R$ {aligned with} a uniform magnetic field ${\bf B} = {B}\hat{z}$,
   where a cylindrical coordinate system $(r, \theta, z)$ is adopted.
{The} equilibrium density 
   is assumed to be a function of $r$ only and of the form
\begin{eqnarray}
\label{eq_rho_profile}
 {\rho}(r)=\left\{
   \begin{array}{ll}
   \rho_{\rm i},    & 0\le r \leq r_{\rm i} = R-l/2, \\
   \rho_{\rm tr}(r),& r_{\rm i} \le r \le r_{\rm e} = R+l/2,\\
   \rho_{\rm e},    & r \ge r_{\rm e} .
   \end{array}
   \right.
\end{eqnarray}
The profile between $[r_{\rm i}, r_{\rm e}]$ is such that
   the equilibrium density $\rho$ decreases continuously from
   the internal value $\rho_{\rm i}$
   to the external one $\rho_{\rm e}$.
The thickness of this transition layer, denoted by $l$, 
    is bounded by {$0$ and $2 R$}.
The former represents the steepest profile of a step-function form,
   whereas the latter corresponds to the least steep case.

\begin{figure}
\centering
\includegraphics[width=0.6\columnwidth]{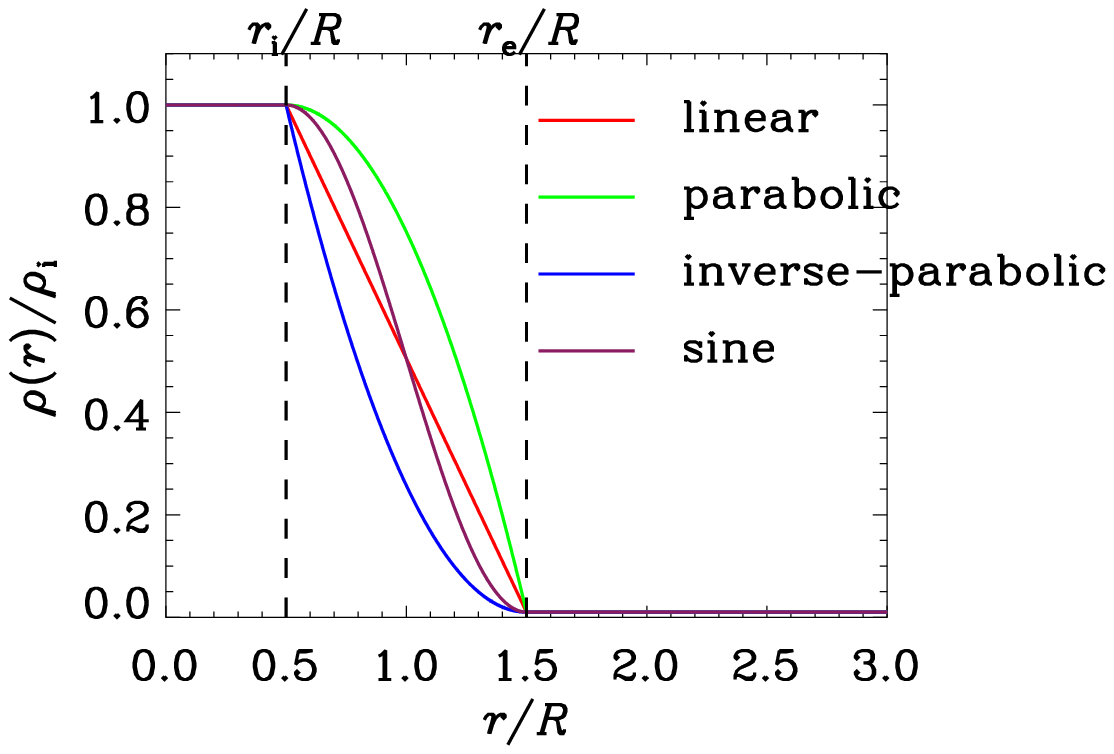}
 \caption{
 Transverse equilibrium density profiles as a function of $r$.
 The profiles differ only in how they are described in a transition layer
     sandwiched between the internal (with a uniform density $\rho_{\rm i}$)
     and external (with a uniform density $\rho_{\rm e}$) parts.
The transition layer is of width $l$, and is located between
     $r_{\rm i}= R-l/2$ and $r_{\rm e}=R+l/2$, with $R$ being the cylinder radius.
Four different choices of the density profiles in the transition layer
     are adopted as labeled, and are given by Eq.~(\ref{eq_rho_tr}).
For illustration purposes, $l$ is chosen to be $R$,
     and $\rho_{\rm i}/\rho_{\rm e}$ is chosen to be $100$.
}
 \label{fig_illus_profile}
\end{figure}

While our analysis is valid for arbitrary prescriptions of $\rho_{\rm tr}$,
   a number of choices have to be made to evaluate quantitatively
   {the effects of equilibrium density profiles}.
To this end, we select the following profiles,
\begin{eqnarray}
\label{eq_rho_tr}
   \rho_{\rm tr}(r)=\left\{
   \begin{array}{ll}
   \rho_{\rm i}-\displaystyle\frac{\rho_{\rm i}-\rho_{\rm e}}{l}\left(r-R+\displaystyle\frac{l}{2}\right),   & {\rm linear},
   \\[0.3cm]
   \rho_{\rm i}-\displaystyle\frac{\rho_{\rm i}-\rho_{\rm e}}{l^2}\left(r-R+\displaystyle\frac{l}{2}\right)^2,& {\rm parabolic},\\[0.3cm]
   \rho_{\rm e}-\displaystyle\frac{\rho_{\rm e}-\rho_{\rm i}}{l^2}\left(r-R-\displaystyle\frac{l}{2}\right)^2,    & {\rm inverse-parabolic}, \\[0.3cm]
   \displaystyle\frac{\rho_{\rm i}}{2}\left[\left(1+
   \displaystyle\frac{\rho_{\rm e}}{\rho_{\rm i}}\right)-\left(1-
   \displaystyle\frac{\rho_{\rm e}}{\rho_{\rm i}}\right)\sin\displaystyle\frac{\pi(r-R)}
   {l}\right],    & {\rm sine}.
   \end{array}
   \right.
\end{eqnarray}
{The profiles} labeled linear, parabolic and sine have been examined in substantial detail
    in the context of standing kink modes~\citep{2013ApJ...777..158S,2014ApJ...781..111S}.
An additional profile, labeled inverse-parabolic, is added to make the list
    more comprehensive {in that} 
    it naturally complements the parabolic one.
Figure \ref{fig_illus_profile} illustrates the
    $r$-dependence of the chosen equilibrium density profiles.
For illustration purposes, {$\rho_{\rm i}/\rho_{\rm e}$
    is chosen to be $100$,
    and $l/R$ is chosen to be unity}.

\subsection{Solutions for Radial Lagrangian Displacement and Total Pressure Perturbation}
\label{sec_sub_xirptot}

Appropriate for the solar corona, we work in the framework of cold (zero-$\beta$) MHD,
   in which case sausage waves do not perturb the $z$-component of the plasma velocity.
Let $\delta v_r$ denote the radial velocity perturbation, and let
$\delta b_r$ and $\delta b_z$
   denote the radial and longitudinal components of the perturbed magnetic field $\delta{\bf b}$, respectively.
{The perturbed total pressure 
   is then $\delta p_{\rm tot} = {\bf B}\cdot\delta{\bf b}/4\pi = B\delta b_z/4\pi$
   given the absence of thermal pressure in
   the zero-$\beta$ limit}.
Fourier-decomposing any perturbed value $\delta f(r, z;t)$ as
\begin{eqnarray}
\label{eq_Fourier_ansatz}
  \delta f(r,z;t)={\rm Re}\left\{\tilde{f}(r)\exp\left[-i\left(\omega t-kz\right)\right]\right\}~,
\end{eqnarray}
   one finds from linearized, ideal, cold MHD equations that
\begin{eqnarray}
\label{eq_xir}
  \frac{1}{r}\left(r\tilde{\xi}_r'\right)'
  +\left(\frac{\omega^2}{v^2_{\rm A}}-k^2-\displaystyle\frac{1}{r^2}\right)\tilde{\xi}_r=0~,
\end{eqnarray}
  where the prime $' = {\rm d}/{\rm d}r$.
In addition, $\tilde{\xi}_r = i\tilde{v}_r/\omega$ is the Fourier
amplitude of the radial Lagrangian displacement,
    and $v_{\rm A}(r)=B/\sqrt{4\pi\rho(r)}$ is the Alfv\'en speed.
{The} Fourier amplitude of the perturbed total pressure is
\begin{eqnarray}
\label{eq_Fourie_ampl}
  \tilde{p}_{\rm tot} = -\frac{{B}^2}{4\pi r}
     \left(r\tilde{\xi}_r\right)' .
\end{eqnarray}

{With azimuthal wavenumber $m$ being $0$, the equations governing linear sausage waves
   are free of singularities,
   making our derivation simpler than in 
   kink mode studies
   where a treatment of singularity is necessary}~\citep{2013ApJ...777..158S}.   
{To be specific, the solutions to Eq.~(\ref{eq_xir}) in the transition layer
   can} be expressed as a regular series expansion
   in $x \equiv r-R$.
{Let $\tilde{\xi}_{\rm tr, 1}$ and $\tilde{\xi}_{\rm tr, 2}$ denote
   two independent solutions},
\begin{eqnarray}
\label{eq_xi1xi2_expansion}
  \tilde{\xi}_{\rm tr, 1}(x) = \sum_{n=0}^\infty a_n x^n~, \hspace{0.2cm}
  \tilde{\xi}_{\rm tr, 2}(x) = \sum_{n=0}^\infty b_nx^n~.
\end{eqnarray}
Without loss of generality, one may choose $[a_0, a_1] = [R, 0]$ and
$[b_0, b_1] = [0, 1]$. Expanding the equilibrium density $\rho$
about $x=0$ as well, one finds that
\begin{eqnarray}
\label{eq_rho_expansion}
  \rho_{\rm tr}(x) = \sum^\infty_{n=0}\rho_n x^n,
\end{eqnarray}
    with $\rho_0 = \rho|_{x=0}$ and
\begin{eqnarray}
\label{eq_rho_coef}
  \rho_n = \frac{1}{n!} \left.\frac{{\rm d}^n\rho(x)}{{\rm d}x^n}\right|_{x=0},
     \hspace{0.2cm} n\ge 1.
\end{eqnarray}
Plugging Eq.~(\ref{eq_xi1xi2_expansion}) into Eq.~(\ref{eq_xir})
with the change of independent variable from
    $r$ to $x$, and then ordering the terms according to powers in $x$, one finds that
\begin{eqnarray}
\label{eq_a_n}
  \chi_2 &=& \frac{-1}{2R^2}\left[R \chi_1+\left(\eta\omega^2R^2\rho_0-k^2R^2-1\right)\chi_0\right]~,  \nonumber \\
  \chi_3 &=& \frac{-1}{6R^2}
           \left[6 R \chi_2 +\left(\eta\omega^2R^2\rho_0-k^2R^2\right)\chi_1 \right.  \nonumber \\
    &&  \left. +\left(\eta\omega^2R^2\rho_1+2\eta\omega^2R\rho_0-2k^2R\right)\chi_0
             \right]~, \\
  \chi_n &=& \frac{-1}{n(n-1)R^2}
       \left\{(n-1)(2n-3)R \chi_{n-1} +\eta\omega^2R^2\sum_{l=0}^{n-2}\rho_{n-l-2} \chi_l \right.   \nonumber \\
    &&  +\left[(n-3)(n-1)-k^2R^2\right]\chi_{n-2} +2\eta\omega^2R\sum\limits_{l=0}^{n-3}\rho_{n-l-3} \chi_l \nonumber \\
    &&   \left. -2k^2R \chi_{n-3}-k^2 \chi_{n-4}+\eta\omega^2\displaystyle\sum\limits_{l=0}^{n-4}\rho_{n-l-4} \chi_l \right\}~, (n\ge4) \nonumber
\end{eqnarray}
    where $\eta = 4\pi/B^2$ and $\chi$ represents either $a$ or $b$.

With Eq.~(\ref{eq_xi1xi2_expansion}) at hand, the solution to Eq.~(\ref{eq_xir})
    can be expressed as
\begin{eqnarray}
\label{eq_xi_solution_entire}
   \tilde{\xi}_r(r)=\left\{
   \begin{array}{ll}
      A_{\rm i}J_1(\mu_{\rm i} r),          & 0         \le r \le r_{\rm i}, \\
      A_1\tilde{\xi}_{\rm tr, 1}(x)+A_2\tilde{\xi}_{\rm tr, 2}(x),      & r_{\rm i} \le r \le r_{\rm e}, \\
      A_{\rm e}H^{(1)}_1(\mu_{\rm e} r),        & r \ge r_{\rm e},
   \end{array} \right.
\end{eqnarray}
    where $A_{\rm i}, A_{\rm e}, A_1$ and $A_2$ are arbitrary constants,
    and $J_n$ and $H_n^{(1)}$ are the $n$-th-order Bessel and Hankel functions of the first kind, respectively (here $n=1$).
In addition, $\mu_{\rm i, e}^2 = \omega^2/v_{\rm Ai, e}^2 - k^2$
with $v_{\rm Ai, e}^2 = {B}^2/(4\pi\rho_{\rm i, e})$. As discussed
in~\citet{1986SoPh..103..277C}, requiring that
    $-\pi/2 < \arg\mu_{\rm i},  \arg\mu_{\rm e} \le \pi/2$ does not exclude any additional independent solution.
Furthermore, expressing the external solution in terms of
$H_n^{(1)}$ permits a unified examination of both trapped and leaky
waves. Indeed, the trapped regime arises when $\arg\mu_{\rm e} =
\pi/2$, in which case one finds that
   $H_1^{(1)}(\mu_{\rm e} r) = -(2/\pi) K_1(\alpha r)$ with $\alpha = \mu_{\rm e}/i$
    being real and positive~\citep[see discussions on page 281 in][]{1986SoPh..103..277C}.
Now the Fourier amplitude for the total pressure perturbation can be
evaluated with Eq.~(\ref{eq_Fourie_ampl}), the results being
\begin{eqnarray}
   \tilde{p}_{\rm tot}(r)=
       \left\{ \begin{array}{ll}
   -\displaystyle\frac{A_{\rm i}{B}^2}{4\pi}\mu_{\rm i} J_0(\mu_{\rm i}r)~,
        & 0 \le r \leq r_{\rm i}, \\[0.3cm]
   -\displaystyle\frac{{B}^2}{4\pi r}
        \left\{ A_1\frac{\rm d}{{\rm d} x}\left[\left(x+R\right) \tilde{\xi}_{\rm tr, 1}(x)\right]
               +A_2\frac{\rm d}{{\rm d} x}\left[\left(x+R\right) \tilde{\xi}_{\rm tr, 2}(x)\right]
        \right\}~,
        & r_{\rm i} \leq  r \leq r_{\rm e}\\[0.3cm]
   -\displaystyle\frac{A_{\rm e}{B}^2}{4\pi}\mu_{\rm e}H^{(1)}_0(\mu_{\rm e}r)~,    & r \ge r_{\rm e},
   \end{array}
   \right.
\end{eqnarray}
The expression for $\tilde{p}_{\rm tot}$ for the ranges $r\le
r_{\rm i}$ and $r\ge  r_{\rm e}$
    can also be independently verified by using an alternative relation between $\tilde{\xi}_r$ and $\tilde{p}_{\rm tot}$,
\begin{eqnarray*}
 \tilde{\xi}_r = \frac{{\rm d}\tilde{p}_{\rm tot}/{\rm d}r}{\rho(\omega^2-k^2 v_{\rm A}^2)}~.
\end{eqnarray*}

\subsection{Dispersion Relation of Sausage Waves}
{The} dispersion relation (DR) governing linear sausage waves can be derived by
     requiring that both $\tilde{\xi}_r$ and $\tilde{p}_{\rm tot}$ be continuous at the interfaces
     $r=r_{\rm i}$ and $r=r_{\rm e}$.
This leads to
\begin{eqnarray*}
 A_{\rm i}J_1(\mu_{\rm i}r_{\rm i}) &=& A_1\tilde{\xi}_{\rm tr,1}(x_{\rm i})
      +A_2\tilde{\xi}_{\rm tr, 2}(x_{\rm i}) , \\
 A_{\rm e}H^{(1)}_1(\mu_{\rm e}r_{\rm e}) &=& A_1\tilde{\xi}_{\rm tr, 1}(x_{\rm e})
      +A_2\tilde{\xi}_{\rm tr, 2}(x_{\rm e}) , \\
 A_{\rm i}\mu_{\rm i} r_{\rm i}J_0(\mu_{\rm i}r_{\rm i}) &=&
   [A_1\tilde{\xi}_{\rm tr,1}(x_{\rm i})+A_2\tilde{\xi}_{\rm tr, 2}(x_{\rm i})]
  +r_{\rm i}[A_1\tilde{\xi}'_{\rm tr,1}(x_{\rm i})+A_2\tilde{\xi}'_{\rm tr, 2}(x_{\rm i})] ,\\
  A_{\rm e}\mu_{\rm e}r_{\rm e}H^{(1)}_0(\mu_{\rm e}r_{\rm e}) &=&
  [A_1\tilde{\xi}_{\rm tr,1}(x_{\rm e})+A_2\tilde{\xi}_{\rm tr, 2}(x_{\rm e})]
  +r_{\rm e}[A_1\tilde{\xi}'_{\rm tr,1}(x_{\rm e})+A_2\tilde{\xi}'_{\rm tr, 2}(x_{\rm e})] ,
\end{eqnarray*}
    where $x_{\rm i}=-l/2$ and $x_{\rm e}=l/2$.
Eliminating $A_{\rm i}$ ($A_{\rm e}$) by dividing the third (fourth) by the first (second) equation, one finds that
\begin{equation}
\label{eq_prep_DR}
\begin{array}{rcl}
&&  \Lambda_1 A_1 + \Lambda_2 A_2 = 0,  \\
&&  \Lambda_3 A_1 + \Lambda_4 A_2 = 0,
\end{array}
\end{equation}
   where the coefficients read
\begin{eqnarray}
\label{eq_Ys}
\begin{array}{rcl}
&& \Lambda_1=\tilde{\xi}_{\rm tr,1}(x_{\rm i})+r_{\rm i}\tilde{\xi}'_{\rm tr,1}(x_{\rm i})-X_{\rm i}\tilde{\xi}_{\rm tr,1}(x_{\rm i}) , \\
&& \Lambda_2=\tilde{\xi}_{\rm tr,2}(x_{\rm i})+r_{\rm i}\tilde{\xi}'_{\rm tr,2}(x_{\rm i})-X_{\rm i}\tilde{\xi}_{\rm tr,2}(x_{\rm i}) , \\
&& \Lambda_3=\tilde{\xi}_{\rm tr,1}(x_{\rm e})+r_{\rm e}\tilde{\xi}'_{\rm tr,1}(x_{\rm e})-X_{\rm e}\tilde{\xi}_{\rm tr,1}(x_{\rm e}) , \\
&& \Lambda_4=\tilde{\xi}_{\rm tr,2}(x_{\rm e})+r_{\rm e}\tilde{\xi}'_{\rm tr,2}(x_{\rm e})-X_{\rm e}\tilde{\xi}_{\rm tr,2}(x_{\rm e}) ,
\end{array}
\end{eqnarray}
with
\begin{equation}
\label{eq_XX}
\begin{array}{rcl}
  X_{\rm i} &=&
    \displaystyle\frac{\mu_{\rm i} r_{\rm i}J_0(\mu_{\rm i}r_{\rm i})}{J_1(\mu_{\rm i}r_{\rm i})} ,\\
[0.3cm]  X_{\rm e} &=&
    \displaystyle\frac{\mu_{\rm e}r_{\rm e}H^{(1)}_0(\mu_{\rm e}r_{\rm e})}{H^{(1)}_1(\mu_{\rm e}r_{\rm e})} .
\end{array}
\end{equation}
Evidently, for Eq.~(\ref{eq_prep_DR}) to allow non-trivial solutions
of $[A_1, A_2]$, one needs to require that
\begin{eqnarray}
\label{eq_DR} \Lambda_1 \Lambda_4 - \Lambda_2 \Lambda_3 = 0 ,
\end{eqnarray}
   which is the DR governing sausage waves in nonuniform loops.

Does Eq.~(\ref{eq_DR}), the DR valid for arbitrary $l/R$,
    recover the well-known result for the step-function profile when $l/R \rightarrow 0$?
Retaining only terms to the 0-th order in $l/R$ and noting that
$r_{\rm i}\approx r_{\rm e} \approx R$,
    one finds {that $\Lambda_{n}~(n=1, \cdots, 4)$}
    simplify to
\begin{eqnarray*}
&&  \Lambda_1=(1-X_{\rm i})a_0+Ra_1 , \\
&&  \Lambda_2=(1-X_{\rm i})b_0+Rb_1 , \\
&&  \Lambda_3=(1-X_{\rm e})a_0+Ra_1 , \\
&&  \Lambda_4=(1-X_{\rm e})b_0+Rb_1 .
\end{eqnarray*}
Substituting these expressions into Eq.~(\ref{eq_DR}), one finds
that
\begin{eqnarray*}
(X_{\rm i}-X_{\rm e})(a_1 b_0-a_0 b_1)=0 .
\end{eqnarray*}
{This leads to $X_{\rm i} = X_{\rm e}$, given that} $a_1 b_0 - a_0 b_1$ is not allowed to be zero
    for $\tilde{\xi}_{\rm tr, 1}$ and $\tilde{\xi}_{\rm tr, 2}$ to be independent.
{In} other words (see Eq.~(\ref{eq_XX}))
\begin{eqnarray}
\label{eq_DR_tophat}
  \frac{\mu_{\rm i}       J_0(\mu_{\rm i} R)}{      J_1(\mu_{\rm i} R)}
 =\frac{\mu_{\rm e} H^{(1)}_0(\mu_{\rm e} R)}{H^{(1)}_1(\mu_{\rm e} R)} ,
\end{eqnarray}
  which is the DR for equilibrium density profiles of a
   step-function form~\citep[e.g.,][]{1986SoPh..103..277C}.

{While the DR is equally applicable to
   propagating waves, we will focus on
   standing modes for which the axial wavenumber $k$ is real, while the angular frequency $\omega$
   is allowed to be complex-valued.
Furthermore, let us
   focus on fundamental standing modes supported by magnetized
   tubes of length $L$}.
In this case, another measure to validate the DR, independent of the eigen-value problem approach,
   is to employ the linearized, time-dependent, cold MHD equations
    to derive an equation governing the transverse velocity perturbation $v_r(r, z, t)$.
In view of the line-tying boundary conditions at the loop ends $z=0$
and $z=L$,
    one may express $v_r(r, z, t)$ as $v(r, t)\sin(kz)$ with $k=\pi/L$,
    yielding~\citep{2012ApJ...761..134N,2015SoPh..tmp..118C}
\begin{equation}
\label{eq_vr}
 \frac{\partial^2 v(r, t)}{\partial t^2}
    =v_{\rm A}^2(r)\left[\displaystyle\frac{\partial^2}{\partial r^2}
    +\displaystyle\frac{1}{r}
    \displaystyle\frac{\partial}{\partial r}-\left(k^2
    +\displaystyle\frac{1}{r^2}\right)\right]v (r, t)~.
\end{equation}
When supplemented with appropriate boundary and initial conditions,
    the signal of $v_r (r, t)$ at
    some arbitrarily chosen distance from the tube axis can be followed.
As demonstrated in~\citet{2012ApJ...761..134N,2015SoPh..tmp..118C},
    after a transitory phase this signal evolves into
    a harmonic (decaying harmonic) form when $k$ is larger (less) than some critical value,
    corresponding to the well-known trapped (leaky) regime.
Numerically fitting the signal with a sinusoidal (exponentially
decaying sinusoidal) function
    then yields the period $P$ ($P$ together with the damping time $\tau$)
    for trapped (leaky) modes.
We also adopt this approach and compare the derived values for $P$ and $\tau$
    with what is found by solving the DR for complex-valued $\omega$ at given real $k$.
As will be shown in Fig.~\ref{fig_Ptau_vs_L}, the two sets of
    independently derived values for $[P, \tau]$ agree remarkably well.
At this {point, it should be remarked} that once a choice for $\rho_{\rm tr}$ is made,
    $P$ and $\tau$ depend only on the combination
    of parameters $[\rho_{\rm i}/\rho_{\rm e}, l/R, L/R]$ when {they} are measured in units of
    the internal Alfv\'en transit time $R/v_{\rm Ai}$.
{Here} we have used $L/R = \pi/(kR)$ in place of the dimensionless axial wavenumber $kR$.

In general, the DR (Eq.~(\ref{eq_DR})) is not analytically tractable
    and {is solved numerically}
    for a given $\rho_{\rm tr}$ profile
    and some given combination of $[\rho_{\rm i}/\rho_{\rm e}, l/R, L/R]$.
To do so requires the infinite series expansion in Eq.(\ref{eq_xi1xi2_expansion}) to be
    truncated by retaining the terms with $n$ up to
    a certain $N$.
A value of $N=101$ is chosen for all the numerical results to be presented,
    and {we have made sure} that using an even larger $N$ does not introduce any appreciable difference.
In addition, we focus only on the lowest order modes, namely those with the simplest radial structure in
    the eigen-functions.
When verifying {these results} with the computations
    from an initial-value-problem perspective,
    we choose not to use a too localized initial perturbation $v(r, t=0)$ (see Eq.(\ref{eq_vr})),
    otherwise higher order modes are introduced to contaminate the $v(r, t)$ signals.

\section{NUMERICAL RESULTS AND THE INVERSION SCHEME}
\label{sec_NumRes}
\begin{figure}
\centering
\includegraphics[width=0.6\columnwidth]{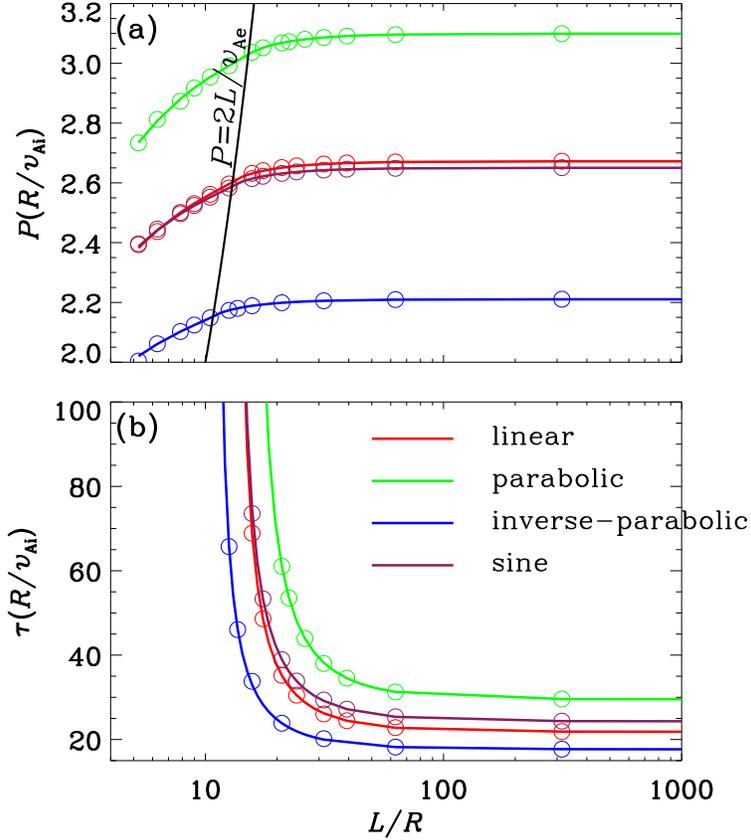}
 \caption{
  (a) Period $P$ and (b) damping time $\tau$ as functions of loop length $L$.
  Four different choices of density profiles are examined as labeled.
  The black line in (a) represents $P=2L/v_{\rm Ae}$ and separates
     the trapped (to its left) from leaky (right) regimes.
  The open circles represent the values for $P$ and $\tau$ obtained by
     solving Eq.~(\ref{eq_vr}) from an initial-value-problem perspective, which
     is independent from the eigen-value-problem approach presented in the text.
  Here the width of the transition layer $l=R$,
     and the density contrast $\rho_{\rm i}/\rho_{\rm e}=100$.
     }
 \label{fig_Ptau_vs_L}
\end{figure}

Let us start with an examination of how {the
     $\rho_{\rm tr}$ profile impacts} the
    dispersive properties of standing sausage modes.
Figure~\ref{fig_Ptau_vs_L} shows the dependence on the length-to-radius ratio $L/R$
   of the period $P$ and damping time $\tau$ for {different choices of
   $\rho_{\rm tr}(r)$ as labeled}.
For illustration purposes, we choose the density contrast $\rho_{\rm i}/\rho_{\rm e}$ to be $100$,
   and choose $l/R$ to be unity.
{The black line in Fig.~\ref{fig_Ptau_vs_L}a, which
   represents $2 L/v_{\rm Ae}$,
   separates the trapped (to its left, where $\tau$ is identically infinite)
   from leaky (to its right) regimes}.
The curves are found by solving the analytical DR (Eq.~(\ref{eq_DR})),
   whereas the open circles are found by solving the corresponding {time-dependent equation}
   (see Eq.~(\ref{eq_vr}) and the associated description).
Evidently, the periods $P$ and damping times $\tau$ obtained from the two independent approaches
   agree with each other remarkably well.
Figure~\ref{fig_Ptau_vs_L} indicates that the overall tendency for $P$ ($\tau$) to increase (decrease)
   with $L/R$ is seen for all the equilibrium density profiles considered.
In particular, regardless of the profiles,
   both $P$ and $\tau$ tend to some asymptotic values at large $L/R$.
However, the choice of equilibrium density profiles
   has a considerable influence on the specific values for $P$ and $\tau$.
{This is particularly true if one compares the results for the parabolic and inverse-parabolic profiles,
   given by the green and blue curves, respectively.
Furthermore, while the periods $P$ for the linear and sine profiles
   differ little for the chosen $l/R$,
   the damping times $\tau$ show a stronger {profile} dependence.
This signifies the importance of using $P$ and $\tau$ in a consistent manner
   when one attempts to deduce how the transverse equilibrium density is structured.}

\begin{figure}
\centering
\includegraphics[width=0.6\columnwidth]{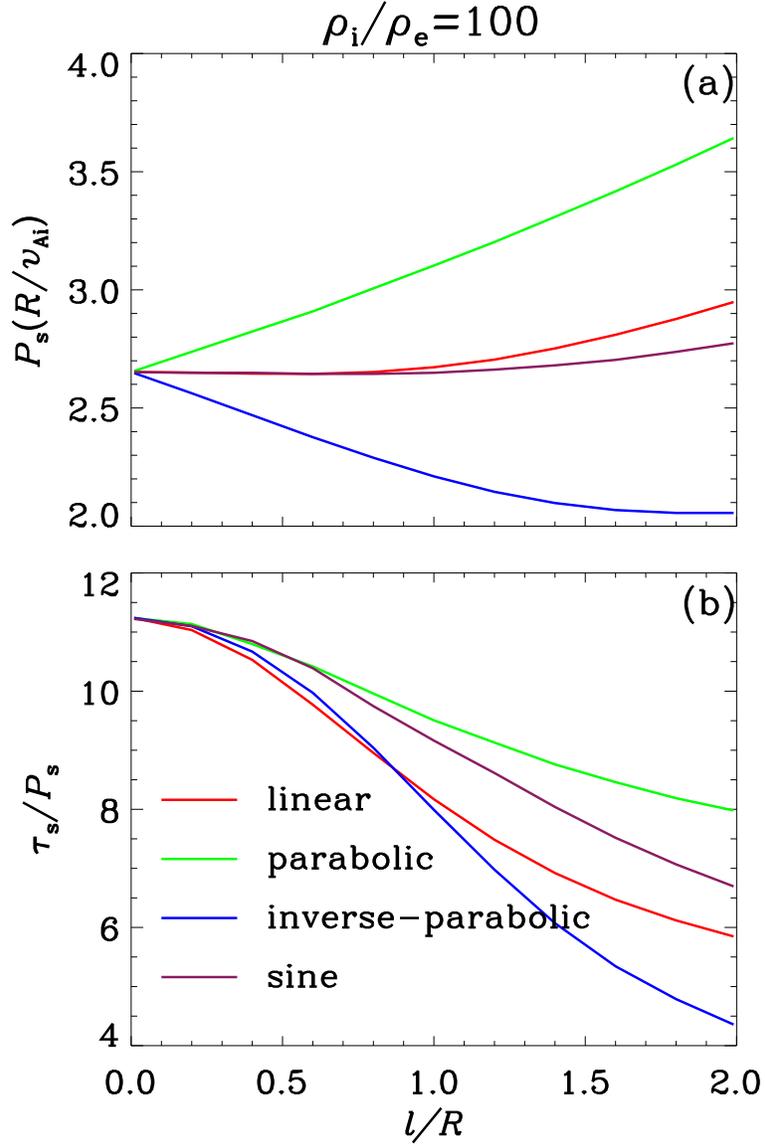}
 \caption{
  Saturation values for (a) period $P_{\rm s}$ and (b) damping time $\tau_{\rm s}$
     as functions of the width of the transition layer $l$.
  These saturation values are attained for sufficiently thin loops.
  Four different choices of the density profiles are examined as labeled.
  Here the density contrast $\rho_{\rm i}/\rho_{\rm e}=100$.
  }
\label{fig_Ptau_vs_loR}
\end{figure}

The effects of {equilibrium} density profile can be better brought out by
   capitalizing on the fact that both $P$ and $\tau$ experience saturation
   for sufficiently thin loops.
Let $P_{\rm s}$ and $\tau_{\rm s}$ denote the saturation values.
Figure~\ref{fig_Ptau_vs_loR} presents how $P_{\rm s}$ and $\tau_{\rm s}/P_{\rm s}$ depend on $l/R$, the width
   of the transition layer in units of loop radius.
The density contrast is chosen to be $100$,
   and different choices of $\rho_{\rm tr}$ are represented by the curves in different colors
   as labeled in Fig.~\ref{fig_Ptau_vs_loR}b.
In addition, in place of $\tau_{\rm s}$, the ratio $\tau_{\rm s}/P_{\rm s}$ is plotted
   since it is a better measure of {signal} quality.
One sees from Figs.~\ref{fig_Ptau_vs_loR}a and \ref{fig_Ptau_vs_loR}b
   that the curves converge at $l/R \rightarrow 0$ as
   expected given that the DR (Eq.~(\ref{eq_DR})) simplifies
   to Eq.~(\ref{eq_DR_tophat}) pertinent to {a top-hat density distribution}.
Figure~\ref{fig_Ptau_vs_loR}a indicates that the $l/R$-dependence of $P_{\rm s}$
   critically depends on how $\rho_{\rm tr}$ is described.
For the parabolic profile, {$P_{\rm s}$ increases monotonically with $l/R$,
   whereas the opposite trend is found for the inverse-parabolic profile.}
When it comes to the linear and sine profiles,
   Fig.~\ref{fig_Ptau_vs_loR}a
   shows that the $l/R$-dependence of $P_{\rm s}$ is not as strong, and
   the difference between the two profiles is discernible only when $l/R\gtrsim 1$.
Moving on to Fig.~\ref{fig_Ptau_vs_loR}b, one sees that
   the ratio $\tau_{\rm s}/P_{\rm s}$ decreases monotonically with $l/R$ for all the profiles,
   meaning that wave leakage plays an increasingly important role
   in attenuating sausage modes when the loop becomes more diffuse.
Reinforcing the impression from Fig.~\ref{fig_Ptau_vs_L}b,
   {one sees that relative to $P_{\rm s}$, $\tau_{\rm s}/P_{\rm s}$ better discriminates
   the equilibrium density profiles.}

\begin{figure}
\centering
\includegraphics[width=0.65\columnwidth]{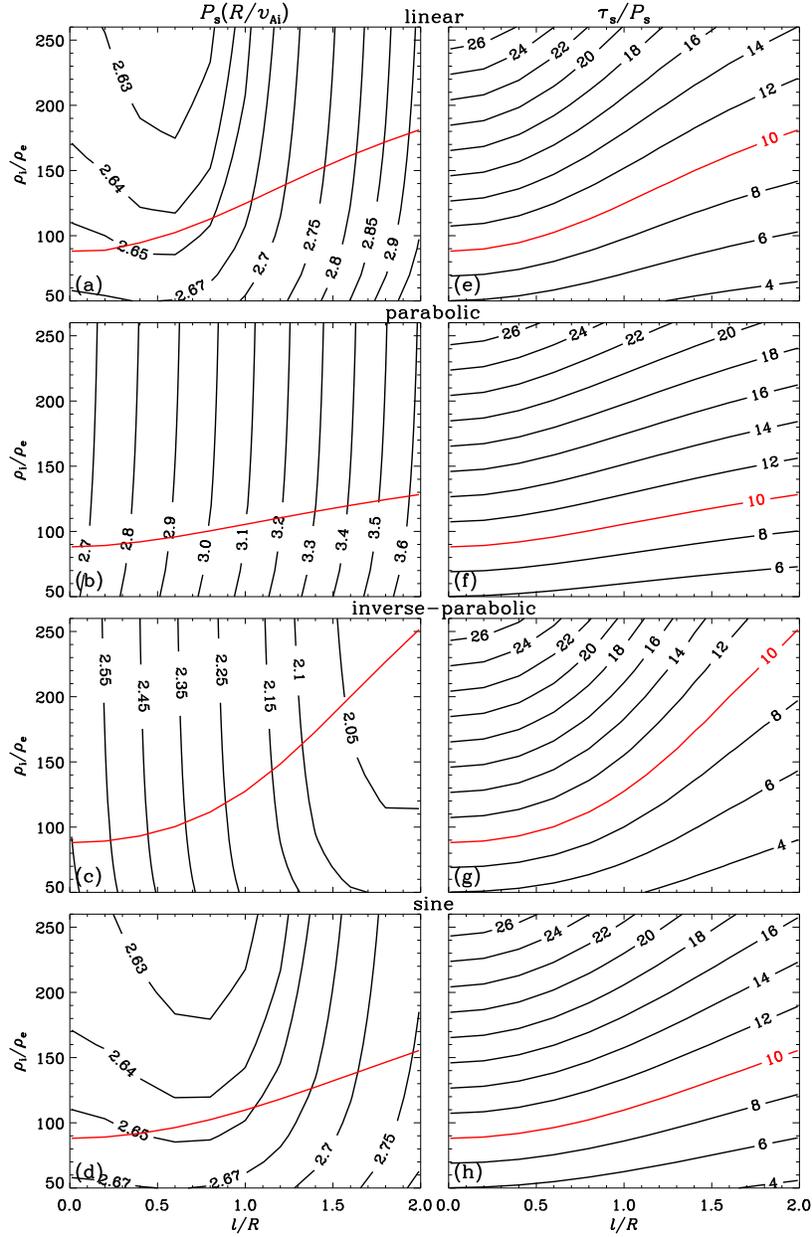}
 \caption{
  Contour plots in the $[l/R, \rho_{\rm i}/\rho_{\rm e}]$ space for
     the saturation values for period $P_{\rm s}$ (the left column)
     and damping-time-to-period ratio $\tau_{\rm s}/P_{\rm s}$ (right).
  These saturation values are attained for sufficiently thin loops.
  Each row represents one of the four different choices of the density profiles as labeled.
  The red curve in each panel represents where $\tau_{\rm s}/P_{\rm s}=10$,
     corresponding to the event reported in~\citet{1973SoPh...32..485M}
  }
\label{fig_contour}
\end{figure}

So far we have fixed the density contrast $\rho_{\rm i}/\rho_{\rm e}$ at $100$.
One naturally asks what happens if $\rho_{\rm i}/\rho_{\rm e}$ is varied?
Figure~\ref{fig_contour} presents the distribution 
    {of $P_{\rm s}$ (the left column) and $\tau_{\rm s}/P_{\rm s}$ (right)}
    in the $[\rho_{\rm i}/\rho_{\rm e}, l/R]$ plane.
Each row represents one of the four density profiles as labeled.
Besides, the red curve represents where $\tau_{\rm s}/P_{\rm s}=10$,
   the value for the QPP event reported in~\citet{1973SoPh...32..485M}.
Examine the left column first.
One sees from Figs.~\ref{fig_contour}b and \ref{fig_contour}c that
   in the parameter range examined, $P_{\rm s}$ tends to increase (decrease) with $l/R$
   at any given $\rho_{\rm i}/\rho_{\rm e}$ when the parabolic (inverse-parabolic) profile
   is chosen.
In contrast, Figs.~\ref{fig_contour}a and \ref{fig_contour}d indicate that
   $P_{\rm s}$ for the linear and sine profiles shows a nonmonotonical
   dependence on $l/R$, {even though this variation is hardly discernible}.
Now consider the right column, from which
    one can see that regardless of the profiles,
    $\tau_{\rm s}/P_{\rm s}$ decreases with increasing $l/R$
    for all the density contrasts examined.
In addition, {the} dependence of
    $\tau_{\rm s}/P_{\rm s}$ on $l/R$ is the strongest for the inverse-parabolic profile,
    and the least strong for the parabolic one.
{The dependence for the linear and sine profiles lies in between,
    with the dependence in the linear case being slightly stronger.}

Conceptually, Fig.~\ref{fig_contour} can be used
    to invert the measured values of 
    the period and damping time of sausage modes, provided that the loops hosting these oscillations
    are sufficiently thin.
Consider the QPP event reported in~\citet{1973SoPh...32..485M} as an example, for which
    $P_{\rm s} = 4.3$~secs and $\tau_{\rm s}/P_{\rm s}=10$.
It then follows that for a given density profile,
    any point along the corresponding red curve in Fig.~\ref{fig_contour} can equally
    reproduce the measured $\tau_{\rm s}/P_{\rm s}$.
After reading any pair of $[\rho_{\rm i}/\rho_{\rm e}, l/R]$,
    one can read from the left column the corresponding value for $P_{\rm s}$ {in units of $R/v_{\rm Ai}$}.
With $P_{\rm s}$ known, one can then deduce $R/v_{\rm Ai}$.
In practice, however, constructing {such a contour plot}
    is not necessary, and one may simply consider the following 3-step inversion scheme.
First, one starts with the dispersion relation for a step-function density profile
    ($l/R=0$), Eq.~(\ref{eq_DR_tophat}),
    to find the value for the density contrast $\rho_{\rm i}/\rho_{\rm e}$
    such that $\tau_{\rm s}/P_{\rm s}$ agrees with the measured value.
Second, with the $\rho_{\rm i}/\rho_{\rm e}$ value for a smaller
$l/R$ as a good guess,
    one can then solve Eq.~(\ref{eq_DR}) to find a new $\rho_{\rm i}/\rho_{\rm e}$
    that yields the measured $\tau_{\rm s}/P_{\rm s}$
    by increasing $l/R$ from $0$ to $2$ consecutively.
Third, with $\rho_{\rm i}/\rho_{\rm e}$ found for all possible
$l/R$,
    one can solve Eq.~(\ref{eq_DR}) for a given pair of $[\rho_{\rm i}/\rho_{\rm e}, l/R]$,
    yielding a value for $P_{\rm s}/(R/v_{\rm Ai})$.
Finding the transverse Alfv\'en transit time $R/v_{\rm Ai}$
    is then straightforward since $P_{\rm s}$ is known.

\begin{figure}
\centering
\includegraphics[width=0.6\columnwidth]{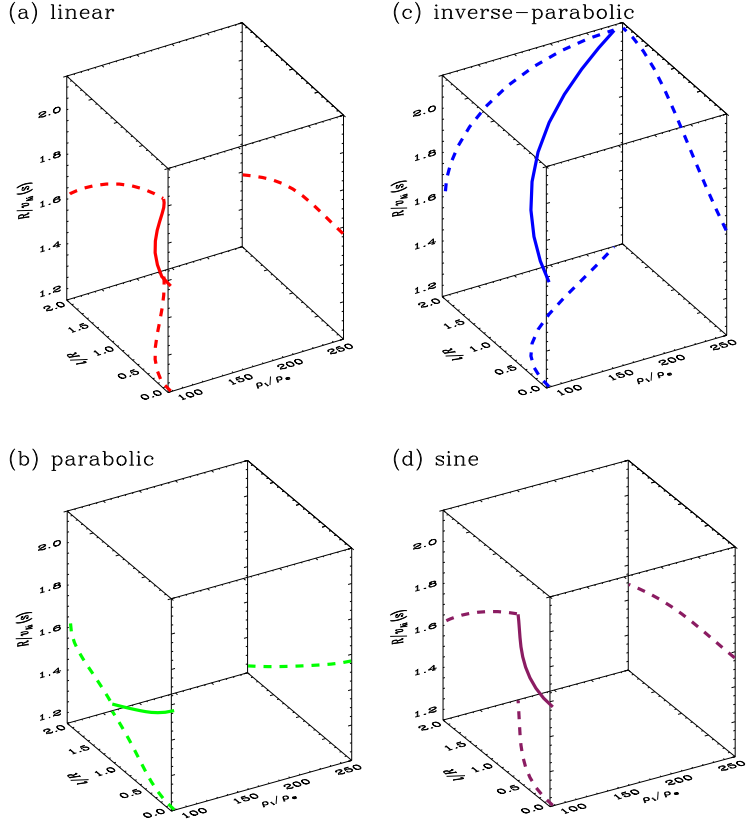}
 \caption{
  Inversion curves (the solid lines) together with their projections (dashed)
     in the three-dimensional space formed by $[R/v_{\rm Ai}, l/R, \rho_{\rm i}/\rho_{\rm e}]$.
  Four choices for the density profiles are examined and given in different panels as labeled.
  All points along an inversion curve are equally compatible with the
     quasi-periodic-pulsation event reported in~\citet{1973SoPh...32..485M} where
      the oscillation period is $4.3$~secs,
      and the damping-time-to-period ratio is $10$.}
\label{fig_3D_results}
\end{figure}

{The product} of the inversion scheme is an inversion curve in the three-dimensional (3D) space
    formed by $R/v_{\rm Ai}$, $l/R$ and $\rho_{\rm i}/\rho_{\rm e}$.
Figure~\ref{fig_3D_results} presents such {curves} (the solid lines)
    and their projections onto various planes (dashed)
    for the examined density profiles, pertinent to the event reported in~\citet{1973SoPh...32..485M}.
To help digest this figure, a number of points {are read from the 
    curves and} presented in Table~\ref{tab_inv}.
{One sees that} 
    among the parameters forming this 3D space,
    $R/v_{\rm Ai}$ is the best constrained.
The biggest (smallest) {value, 
    $2.13$ ($1.18$) secs, is}
    found for the inverse-parabolic (parabolic) profile when $l/R \rightarrow 2$.
{In other words}, the biggest value exceeds the smallest one by
    only $79.7\%$.
As to the density contrast, the biggest value
    ($251.4$ found for the inverse-parabolic profile when $l/R \to 2$)
    is larger than the smallest one ($88.1$ when $l/R\rightarrow 0$)
    by $185\%$.
The least constrained parameter is $l/R$, {with
    any value in the allowed range from $0$ to $2$
    being possible}.

\section{FURTHER DEVELOPMENT OF THE INVERSION SCHEME}
\label{sec_spatial_res}
Before proceeding, let us first recap the key points in the scheme for inverting the measured period $P$ and damping time $\tau$
    of sausage modes.
From the outset, we have assumed that only $P$ and $\tau$ are known,
    whereas we have no information on either the geometric parameters ($R$, $L$, and $l$)
    or the physical parameters ($v_{\rm Ai}$, $\rho_{\rm i}$, and $\rho_{\rm e}$).
On top of that, the specific density profile is also assumed to be unknown.
This happens when one has only spatially unresolved observations~\citep[see e.g., the majority of
    the events compiled in][Table 1]{2004ApJ...600..458A}.
In this case, the dispersion relation, Eq.~(\ref{eq_DR}), suggests that in general the periods and damping times can be
    formally expressed as
\begin{eqnarray}
\label{eq_FG_general}
\begin{array}{rcl}
&& P_{\rm saus} = \displaystyle\frac{R}{v_{\rm Ai}}
     F_{\rm saus}\left(\displaystyle\frac{L}{R}, \displaystyle\frac{l}{R}, \displaystyle\frac{\rho_{\rm i}}{\rho_{\rm e}} \right), \\ [0.4cm] &&
\displaystyle\frac{\tau_{\rm saus}}{P_{\rm saus}} =
     G_{\rm saus}\left(\displaystyle\frac{L}{R}, \displaystyle\frac{l}{R}, \displaystyle\frac{\rho_{\rm i}}{\rho_{\rm e}} \right).
\end{array}
\end{eqnarray}
With only two measured values available,
    any point on a two-dimensional (2D) surface in the four-dimensional (4D) space formed by $[R/v_{\rm Ai}, L/R, l/R, \rho_{\rm i}/\rho_{\rm e}]$
    can reproduce the measurements, even if one is allowed to prescribe a density profile.
{If the sausage modes}
    are in the trapped regime in the sense that the QPPs do not show temporal damping,
    then the situation becomes even less desired since now the restriction from $\tau$ is no longer available.
This complexity can be alleviated if the condition $L/R \gg 1$ holds for the flare loops in question since
    $L/R$ no longer appears such that Eq.~(\ref{eq_FG}) is restored.
One then finds a curve in the $[R/v_{\rm Ai}, l/R, \rho_{\rm i}/\rho_{\rm e}]$ space as shown by Fig.~\ref{fig_3D_results}.

The situation improves if the QPP events are spatially resolved,
    since the looplength ($L$) and the outer interface of the loops ($r_{\rm e} = R+l/2$)
    can be considered known.
In this case $L/R$ and $l/R$ are no longer independent but are related by
\begin{eqnarray}
\label{eq_LR_lR}
    \frac{L}{R} = \frac{L}{r_{\rm e}}\left(1+\frac{l}{2 R}\right) ,
\end{eqnarray}
    where we have used the relation
\begin{eqnarray}
\label{eq_R_loR}
    R = \frac{2 r_{\rm e}}{2 + l/R}.
\end{eqnarray}
{Now the situation is similar to what Eq.~(\ref{eq_FG}) implies}:
    if the sausage mode is a trapped (leaky) one, then a 2D surface (1D curve)
    can be deduced in the $[R/v_{\rm Ai}, l/R, \rho_{\rm i}/\rho_{\rm e}]$ space.
It is just that now for a point on this 2D surface or along the 1D curve,
    one can further deduce $v_{\rm Ai}$ in view of Eq.~(\ref{eq_R_loR}).

Something fascinating {happens} if one observes a spatially resolved QPP event hosting
    more than just a sausage mode.
{For illustration purposes, we consider the situation where a temporally decaying kink mode is involved,
    and its damping can be attributed to resonant absorption}. 
In this case, the period and damping time for the kink mode can be formally expressed by
\begin{eqnarray}
\label{eq_FG_kink_general}
\begin{array}{rcl}
&& P_{\rm kink} = \displaystyle\frac{L}{v_{\rm Ai}}
     F_{\rm kink}\left(\displaystyle\frac{l}{R}, \displaystyle\frac{\rho_{\rm i}}{\rho_{\rm e}} \right), \\ [0.4cm] &&
\displaystyle\frac{\tau_{\rm kink}}{P_{\rm kink}} =
     G_{\rm kink}\left(\displaystyle\frac{l}{R}, \displaystyle\frac{\rho_{\rm i}}{\rho_{\rm e}} \right).
\end{array}
\end{eqnarray}
Note that while $L/R$ in principle can be incorporated into Eq.~(\ref{eq_FG_kink_general}),
    in reality there is no need to do so since the corrections to $P_{\rm kink}$ and $\tau_{\rm kink}$
    due to finite $L/R$ are of the order $(R/L)^2$~\citep{2004ApJ...606.1223V, 2008A&A...484..851G}.
Even for relatively thick flare loops,
    $R/L$ is of the order $0.1$ and these corrections amount to only a few percent.
Note further that $F_{\rm kink}$ and $G_{\rm kink}$ have been extensively studied,
    and a graphical representation can be found in Fig.~1 of~\citet{2014ApJ...781..111S}.
With $F_{\rm kink}$, $G_{\rm kink}$, $F_{\rm saus}$, and $G_{\rm saus}$ known,
    it is then possible to fully constrain the unknowns
    $[l, R, v_{\rm Ai}, \rho_{\rm i}/\rho_{\rm e}]$,
    if one assumes a density profile.
In fact, the measured values of $[P_{\rm kink}, \tau_{\rm kink}, P_{\rm saus}, \tau_{\rm saus}]$
    are more than sufficient: $l$ and $R$ are not independent but are related through Eq.~(\ref{eq_R_loR}).
{This suggests that only three expressions contained in Eqs.~(\ref{eq_FG_general}) and (\ref{eq_FG_kink_general})
    are needed}.
In practice, we consider the expression for the kink mode period as the redundant one.
Now the inversion procedure is rather straightforward.
In view of the relation~(\ref{eq_LR_lR}),
    the two expressions for the damping-time-to-period ratio in Eqs.~(\ref{eq_FG_general})
    and Eqs.~(\ref{eq_FG_kink_general})
    contain only two unknowns, namely $l/R$ and $\rho_{\rm i}/\rho_{\rm e}$.
With both $\tau_{\rm saus}/P_{\rm saus}$ and $\tau_{\rm kink}/P_{\rm kink}$ available,
    one can deduce a unique pair of $[l/R, \rho_{\rm i}/\rho_{\rm e}]$.
The loop radius $R$ then follows from Eq.~(\ref{eq_R_loR}), which
    then enables one to evaluate $v_{\rm Ai}$ with the first expression in Eq.~(\ref{eq_FG_general}).
Finally, as a safety check, one can proceed to evaluate, with the first expression in Eq.~(\ref{eq_FG_kink_general}),
    the theoretically expected kink mode period $P_{\rm kink, theory}$.
The deviation of $P_{\rm kink, theory}$ from the measured one
    then allows to say a few words on how safe it is to identify the oscillation signals
    with some particular modes.

As an illustration of the aforementioned inversion procedure, let us consider
    the QPP event in microwave emissions measured with the Nobeyama Radioheliograph (NoRH)
    on 14 May 2013~\citep{2015A&A...574A..53K}.
Lucky enough, it is likely that this event contains
    a fundamental kink mode with $P_{\rm kink} = 100$~secs and $\tau_{\rm kink}/P_{\rm kink} = 2.5$,
    in addition to a fundamental sausage mode with $P_{\rm saus} = 15$~secs
    and $\tau_{\rm saus}/P_{\rm saus} = 6$.
Assuming that the apparent width measured therein corresponds to $2 r_{\rm e}$, one finds that
    $r_{\rm e}=4\times 10^3$~km and $L=4\times 10^4$~km, meaning that $L/r_{\rm e} = 10$.
{Let} us assume that
    wave leakage is responsible for damping the sausage mode,
    and resonant absorption is responsible for damping the kink one.
Furthermore, let us assume that the sine profile best describes the equilibrium density distribution.
{In this case, the analytical expressions
    obtained in the thin-tube-thin-boundary approximation
    are accurate to within $\sim 25\%$} \citep[e.g.,][]{2004ApJ...606.1223V,2014ApJ...781..111S}.
This enables us to illustrate our inversion procedure without resorting to
    a fully numerical solver to establish $F_{\rm kink}$ and $G_{\rm kink}$.
{Now} the formal expressions given by Eq.~(\ref{eq_FG_kink_general}) can
    be replaced with~\citep[e.g.,][]{2008A&A...484..851G}
\begin{eqnarray}
\label{eq_FG_kink_sine_TTTB}
\begin{array}{rcl}
&& P_{\rm kink} = \displaystyle\frac{L}{v_{\rm Ai}}
     \sqrt{\frac{2(1+\rho_{\rm i}/\rho_{\rm e})}{\rho_{\rm i}/\rho_{\rm e}}}, \\ [0.4cm] &&
\displaystyle\frac{\tau_{\rm kink}}{P_{\rm kink}} =
     \frac{2}{\pi}\frac{\rho_{\rm i}/\rho_{\rm e}+1}{\rho_{\rm i}/\rho_{\rm e}-1}
     \frac{1}{l/R} .
\end{array}
\end{eqnarray}
Following the outlined inversion procedure, we find that
    $l/R = 0.272$ and $\rho_{\rm i}/\rho_{\rm e} = 29.8$ as constrained by the ratios $\tau/P$.
It then follows that $R=3.52 \times 10^3$~km, from which
    one can deduce that $v_{\rm Ai} = 623$~km~s$^{-1}$.
Finally, $P_{\rm kink, theory}$ is found to be $92.2$~secs, which agrees with
    the measured value of $100$~secs to within $8\%$.
This safety check lends support to the interpretation of the two modes in terms of
    fundamental kink and sausage modes as done by~\citet{2015A&A...574A..53K}.
Besides, the deduced Alfv\'en speed $v_{\rm Ai}$ and density contrast $\rho_{\rm i}/\rho_{\rm e}$
    both seem reasonable.
On top of that, there is no need for one to worry too much about the accuracy of Eq.~(\ref{eq_FG_kink_sine_TTTB})
    describing the kink mode:
    at the deduced $[l/R, \rho_{\rm i}/\rho_{\rm e}]$, this equation yields
    $P$ and $\tau/P$ to an accuracy better than $1\%$ and $\sim 6\%$, respectively
    (see Fig.~1 in \citeauthor{2014ApJ...781..111S}~\citeyear{2014ApJ...781..111S}).

\section{CONCLUSIONS}
\label{sec_conc}

How {plasma density} is structured across various magnetic structures in the solar corona
    remains largely unknown.
It has been a common practice to deduce this key information by employing magneto-seismological techniques
    that invert the measured period $P$ and damping time $\tau$
    of standing kink modes collectively supported by
    a magnetic structure~\citep[e.g.,][]{2008A&A...484..851G, 2014ApJ...781..111S}.
In contrast, while {quasi-periodic pulsations} (QPPs) in the lightcurves of solar flares
    are often attributed to standing sausage modes in flare loops
    and therefore can also offer {the associated period and damping time},
    a scheme is missing for inverting these two measurements to deduce the information
    on the density distribution transverse to flare loops.
The primary aim of this study has been to construct such a scheme.
To this end, we worked in the framework of cold (zero-$\beta$) MHD and modeled
    flare loops as straight cylinders with a
    transverse density profile characterized by a transition layer sandwiched between
    a uniform cord and a uniform external medium.
An analytical dispersion relation (DR) governing linear sausage waves,
    Eq.~(\ref{eq_DR}),
    was derived by solving the perturbation equations in terms of
    a regular series expansion in the transition layer.
This DR, valid for arbitrary choices of the density profile in the transition layer,
    then enabled us to examine the effects of density structuring on the {periods and damping times}
    of sausage modes, thereby facilitating the construction of the inversion scheme.

In general, we found that $P$ and $\tau$ of sausage modes depend on a combination of parameters
    $[R/v_{\rm Ai}, L/R, l/R, \rho_{\rm i}/\rho_{\rm e}]$ as formally expressed by Eq.~(\ref{eq_FG_general}),
    where the functions $F_{\rm saus}$ and $G_{\rm saus}$ are a product of the DR.
Here $L$ and $R$ denote the looplength and loop radius, respectively.
Furthermore, $l$ is the width of the transition layer, $v_{\rm Ai}$ is the
    Alfv\'en speed in the cord,
    and $\rho_{\rm i}/\rho_{\rm e}$ is the density contrast between the loop and its surrounding fluid.
We showed that for the density profiles examined, both $P$ and $\tau$
    experience saturation for sufficiently large $L/R$ when the rest of the four parameters are fixed.
The choice of the transverse density profile was found to have a considerable influence
    on $P$ and $\tau$, their dependence on $l/R$ in particular.

Our inversion scheme can find applications to both spatially unresolved and resolved QPP events.
For spatially unresolved ones, we showed that the best one can do is to deduce a 1D curve
    in the $[R/v_{\rm Ai}, l/R, \rho_{\rm i}/\rho_{\rm e}]$ space for a prescribed
    density profile.
This happens if the QPPs in question experience temporal damping, and if the flare loops hosting them
    {can be assumed to be sufficiently thin}.
When applied to a QPP event reported by~\citet{1973SoPh...32..485M}, this inversion technique indicated
    that the transverse Alfv\'en transit time $R/v_{\rm Ai}$ is the best constrained, varying by
    a factor of $~80\%$ if the uncertainties in specifying the density profile are taken into account.
The density contrast $\rho_{\rm i}/\rho_{\rm e}$ is less well constrained, with the largest deduced value
    exceeding the smallest one by a factor of $1.85$.
The least constrained is the transverse density lengthscale in units of loop radius ($l/R$),
    any value in the allowed range $(0, 2)$ can be equally possible to reproduce the measurements.

When it comes to spatially resolved events, the geometric parameters $L$ and $R+l/2$ are additional
    constraints, on top of the measured values for $P$ and $\tau$.
Even though in this case one cannot assume $L/R \gg 1$ a priori,
    it is possible to deduce a 1D curve in the $[v_{\rm Ai}, l, \rho_{\rm i}/\rho_{\rm e}]$
    space for a chosen density profile since $R$ is expressible in terms of $l/R$.
If a spatially resolved QPP event comprises more than just one sausage mode, then it is possible
    to deduce the full information on $[l, R, v_{\rm Ai}, \rho_{\rm i}/\rho_{\rm e}]$.
In this case, the inversion problem becomes an over-determined one.
We illustrated this fascinating application with the QPP event reported by~\citet{2015A&A...574A..53K}
    where a fundamental kink and a fundamental sausage mode were suggested to co-exist
    and both experience temporal damping.
Attributing the temporal damping of the kink mode to resonant absorption,
    and that of the sausage mode to wave leakage,
    we were able to deduce the full set of $[l, R, v_{\rm Ai}, \rho_{\rm i}/\rho_{\rm e}]$
    by using Eqs.~(\ref{eq_FG_general}) and (\ref{eq_FG_kink_sine_TTTB})
    and assuming a sine profile for the density distribution.
One redundant equation, here taken as the expression for the kink mode period,
    can then allow a safety check on, say, whether it is reasonable to interpret the signals in the QPP event
    as the aforementioned modes.
For this particular event, our results demonstrated that the interpretation provided in~\citet{2015A&A...574A..53K}
    is reasonable, not only because the deduced parameters seem realistic,
    but also because the theoretical prediction for the kink mode period agrees
    with the measured one remarkably well.

Our scheme nonetheless has a number of limitations.
First, we have assumed that the temporal damping of sausage modes is due to wave leakage, an ideal MHD process.
In reality, non-ideal mechanisms like electron heat conduction and ion viscosity can provide additional channels for
   damping sausage modes.
While these non-ideal processes were shown by \citet{2007AstL...33..706K}
   to be unlikely the cause for the temporal damping in the QPP event
   reported by~\citet{1973SoPh...32..485M},
   their importance has yet to be assessed for
   the event reported by \citet{2015A&A...574A..53K}.
Second, working in the cold MHD, we have not taken into account the possible effects due to
   finite plasma beta, which may be of the order unity in flare loops.
However, the corrections due to finite beta seem marginal~\citep{2009A&A...503..569I}.
Third, the longitudinal variation in neither the plasma density nor the magnetic field strength
   has been considered, even though the corrections due to this variation
   are unlikely to be significant~\citep{2009A&A...494.1119P}.
Fourth, the density inhomogeneity in flare loops was assumed to be in a monolithic form, whereas
   in reality these loops may be multi-stranded.
While the fine structuring in the form of randomly distributed concentric shells
   is found to have a far less significant influence than the monolithic
   component of the density distribution
   (\citeauthor{2015SoPh..tmp..118C}~\citeyear{2015SoPh..tmp..118C}, see also
   \citeauthor{2007SoPh..246..165P}~\citeyear{2007SoPh..246..165P}),
   there is a need to rigorously assess the effects due to fine structuring in the form of
   randomly distributed strands.

\acknowledgments
{We thank the referee for the constructive comments, 
   which helped improve this manuscript substantially.}
This work is supported by the
    National Natural Science Foundation of China (41174154, 41274176, and 41474149),
    and by the Provincial Natural Science Foundation of Shandong via Grant JQ201212.

\bibliographystyle{apj}
\bibliography{chen}

\IfFileExists{\jobname.bbl}{} {\typeout{}
\typeout{****************************************************}
\typeout{****************************************************}
\typeout{** Please run "bibtex \jobname" to obtain} \typeout{**
the bibliography and then re-run LaTeX} \typeout{** twice to fix
the references !}
\typeout{****************************************************}
\typeout{****************************************************}
\typeout{}}

\pagebreak

\pagebreak

\pagebreak

\pagebreak

\pagebreak

\clearpage
\begin{table}[htbp]
\begin{tabular}{l|cc|cc|cc|cc}
  \hline\hline
\multirow{2}{*}{$l/R$} &  \multicolumn{2}{|c|}{linear} &
\multicolumn{2}{|c}{parabolic}
&  \multicolumn{2}{|c|}{inverse-parabolic} & \multicolumn{2}{|c}{sine} \\
\cline{2-9} &  $\rho_{\rm i}/\rho_{\rm e}$ & $R/v_{\rm Ai}$~(sec) &
$\rho_{\rm i}/\rho_{\rm e}$ & $R/v_{\rm Ai}$~(sec) &  $\rho_{\rm
i}/\rho_{\rm e}$ & $R/v_{\rm Ai}$~(sec)
&  $\rho_{\rm i}/\rho_{\rm e}$ & $R/v_{\rm Ai}$~(sec) \\
[0.2cm]\hline
 0.01 &  88.1   & 1.62   & 88.2  & 1.62  & 88.1  & 1.62   & 88.1    & 1.62  \\
0.2   &  88.9   & 1.62   & 89.2  & 1.57  & 89.3  & 1.68   & 89.1    & 1.62  \\
0.4   &  94.6   & 1.63   & 92    & 1.52  & 93.2  & 1.74   & 91.9    & 1.62  \\
0.6   &  102.4  & 1.63   & 95.8  & 1.48  & 100.4 & 1.81   & 96.4    & 1.63  \\
0.8   &  112.75 & 1.62   & 100.4 & 1.43  & 111.5 & 1.88   & 102.4   & 1.63  \\
1.0   &  124.7  & 1.61   & 105.4 & 1.39  & 127.5 & 1.95   & 109.8   & 1.62  \\
1.2   &  137.4  & 1.6    & 110.5 & 1.35  & 148.4 & 2.02   & 118.2   & 1.62  \\
1.4   &  149.8  & 1.57   & 115.4 & 1.3   & 173.2 & 2.07   & 127.4   & 1.61  \\
1.6   &  161.5  & 1.54   & 120   & 1.26  & 200.1 & 2.1    & 136.9   & 1.6   \\
1.8   &  172    & 1.51   & 124.4 & 1.22  & 226.8 & 2.12   & 146.4   & 1.58  \\
1.99  &  181.1  & 1.47   & 128.2 & 1.18  & 251.4 & 2.13   & 155.4   & 1.56  \\
  \hline
\end{tabular}
\caption{The inverted values for the transverse density length scale
in units of loop radius ($l/R$),
    density contrast $\rho_{\rm i}/\rho_{\rm e}$ and the transverse Alfv\'en transit time $R/v_{\rm Ai}$.
  This inversion is made for the quasi-periodic-pulsation event reported in~\citet{1973SoPh...32..485M},
    assuming that the associated flare loop is sufficiently thin.
  For this event, the oscillation period is $4.3$ secs, and the damping-time-to-period ratio is $10$.
  }
\label{tab_inv}
\end{table}

\end{document}